\RequirePackage{fix-cm}
\documentclass[smallcondensed,smallextended,twocolumn]{svjour3}
\smartqed  
\usepackage{latexsym,amsmath,amssymb,amsfonts,mathbbol,graphicx,color}
\begin{document}

\title{Entanglement Evolution in a Five Qubit Error Correction Code
}


\author{Yaakov S. Weinstein}


\institute{Quantum Information Science Group, {\sc Mitre}, 260 Industrial Way West, Eatontown, NJ 07224, USA
              \email{weinstein@mitre.org}           
}

\date{Received: date / Accepted: date}

\maketitle

\begin{abstract}
In this paper I explore the entanglement evolution of qubits that are part of 
a five qubit quantum error correction code subject to various decohering environments. 
Specifically, I look for possible parallels between the entanglement 
degradation and the fidelity of the logical qubit of quantum information stored in 
the physical qubits. In addition, I note the possible exhibition of entanglement 
sudden death (ESD) due to decoherence and question whether ESD is actually a roadblock
to successful quantum computation. 
\end{abstract}

\section{Introduction}

It is currently accepted that the leading obstacle on the path towards practical quantum 
computation is the inevitibility of decoherence which stems from unwanted interactions 
between the system of interest and its environment \cite{book}. A consequence of decoherence
is the degradation of entanglement between subsystems (such as qubits). Entanglement between
qubits is thought to be necessary for proper operation of a quantum computer. 
Though the need to fight the effects of decoherence has long been realized, recent 
research has suggested that the these effects may be worse than 
originally thought. The coherence of a system may approach zero asymptotically 
due to unwanted interactions with the environment but the entanglement 
may completely disappear in a finite amount of time \cite{D,H,YE1,YE2}. This comparatively 
sudden disappearance of entanglement is termed entanglement sudden death (ESD).
Recent theoretical studies have been devoted to understanding this phenomenon in bi- and 
multi-partite systems \cite{ACCAD,LRLSR,YYE1,YYE2,BYW,YSWx} and there have been a number of 
experimental studies of this phenomenon as well \cite{expt1,expt2,expt3}. Importantly, there 
has been a call to formulate methods of avoiding the ESD phenomenon \cite{YE3} so as to 
ensure the viability of quantum computation.

Yet, it is not at all clear that quantum computer architects should be 
concerned with ESD in any way more than they are concerned with general issues of 
decoherence. A series of recent papers has shown that the onset of ESD has no 
singular effect on the accuracy of certain quantum protocols. ESD causes neither a dramatic 
drop in protocol accuracy, as measured for example by the fidelity, nor any substantive 
change in protocol behavior. Instead, for protocols such as a three qubit error correction code 
\cite{YSW1}, a cluster-based single qubit rotation \cite{YSW2}, and a decoherence free 
subspace \cite{YSW3}, ESD is a non-descript byproduct of decoherence. 

In this paper I continue to explore entanglement evolution in a system implementing 
a quantum protocol in a decohering environment. Specifically, I look at the entanglement 
between the qubits of a five-qubit quantum error correction (QEC) code \cite{5Q,5Q2}. In 
contrast to the previously explored three qubit QEC code explored in \cite{YSW1}, the five 
qubit QEC code can fully protect one qubit of quantum information from all possible single qubit errors. 
The five physical qubits of the QEC code are usually entangled. In fact, it would not be possible 
to design a five qubit QEC protocol that fully protects one qubit of quantum information 
without entanglement between the constituent physical qubits. Based on this I address the following:
how does the loss of entanglement affect the ability of the code to protect the quantum information?
To address this question I compare the degradation of entanglement as a function of decoherence strength 
to the fidelity of the stored logical qubit. We will see that for certain initial states the fidelity 
of the stored information remains high despite significant loss of entanglement and, in general, 
the decay of stored quantum information fidelity does not strongly correlate with the entanglement behavior. 
Finally, I show that the negative effects of decoherence render the QEC code useless well before a complete 
loss of entanglement in the system. In fact, in most cases explored below ESD does not occur before 
complete decoherence. This lack of correlation suggests that entanglement $per s\acute{e}$ is not what 
drives this QEC protocol. Rather, most states in Hilbert space are entangled and the specific states 
utilized by the five qubit QEC are typical in that regard.

\subsection{The Five Qubit Error Correction Code}

For our study of the entanglement evolution, I start with an unencoded single qubit in the state 
$|\psi_u\rangle = \cos\alpha|0\rangle+e^{-i\beta}\sin\alpha|1\rangle$. I assume 
perfect encoding of this qubit of quantum information into five qubits whose 
state after encoding is $|\psi_L\rangle = \cos\alpha|0_L\rangle+e^{-i\beta}\sin\alpha|1_L\rangle$. 
There are a number of formulations of five qubit QEC codes that can fully protect one qubit of 
quantum information. Here I use the formulation of \cite{5Q} with logical $|0\rangle$
and $|1\rangle$:
\begin{eqnarray}
|0_L\rangle &=& |00000\rangle-|01111\rangle+|10011\rangle+|11100\rangle \nonumber\\
		&+& |00110\rangle+|01001\rangle+|10101\rangle+|11010\rangle \nonumber\\ \nonumber\\
|1_L\rangle &=& -|11111\rangle+|10000\rangle+|01100\rangle+|00011\rangle \nonumber\\
		&+& |11001\rangle+|10110\rangle+|01010\rangle+|00101\rangle 
\end{eqnarray}
The qubits are placed in a decohering environment where the five physical qubits are subject to
decoherence of strength $\delta$. The error syndrome is determined by measuring qubits 1, 2, 
4, and 5 and the appropriate recovery operation is applied. If $\delta$ is small the syndrome 
measurement will project the qubits into a state where, up to order $\delta^2$, at most one error 
has occurred. The error will be corrected by the recovery operation. The exact output state, however,
will depend on the outcome of the syndrome measurement. Thus, to quantify the fidelity of 
the stored quantum information I will use as the final single qubit state, $\rho_f(\alpha,\beta,\delta)$, 
the mixed state weighted average of all 16 possible syndrome measurement outcomes (after 
application of the appropriate recovery operation). 

\subsection{Entanglement and Accuracy Measures}

To quantify and monitor entanglement between physical qubits within the QEC code
as they are subject to decoherence I use an entanglement measure known as the negativity, 
$N$ defined as the most negative eigenvalue of the parital transpose of the system density 
matrix \cite{neg}. There are a number of inequivalent forms of the negativity for any 
multi-qubit system: the partial transpose may be taken with respect to any single qubit, $N_j$, 
or the partial transpose may be taken with respect to any two qubits, $N_{j,k}$. Note that 
a zero value of all negativities does not guarantee separability of the state though it does
mean that any entanglement that is present is not distillable. 

As an accuracy measure for the single qubit of quantum information stored in the 
QEC code I use the fidelity,
\begin{equation}
F(\alpha,\beta,\delta) = \langle\psi_u|\rho_f|\psi_u\rangle,
\end{equation}
which is a measure of how well the QEC code has protected the single qubit of logical
information. Here we are especially interested in comparing the fidelity of the 
qubit of stored information with the amount of, and the degradation of, entanglement 
in the system before syndrome measurement.

\section{Decoherence Models}

In each of the next three subsections I explore different decohering environments in which 
the five qubits of the QEC code are placed. As we shall see, each of the environments affects
the system very differently, both with respect to the fidelity of the stored information and 
with respect to the entanglement evolution. The three decoherence models are the independent 
qubit phase damping, amplitude damping, and depolarizing environments.

\subsection{Phase Damping}

We first look at the entanglement evolution of the five qubit system with no
interaction between the qubits, in an independent qubit dephasing environment. 
This environment is fully described by the Kraus operators
\begin{equation}
K_1 = \left(
\begin{array}{cc}
1 & 0 \\
0 & \sqrt{1-\delta} \\
\end{array}
\right); \;\;\;\;
K_2 = \left(
\begin{array}{cc}
0 & 0 \\
0 & \sqrt{\delta} \\
\end{array}
\right),
\end{equation} 
where the dephasing parameter $\delta$ can also be written in a time-dependent fashion such as
$\delta = 1-\exp(-\kappa t)$ for some suitable decay constant $\kappa$ and time $t$.

\begin{center}
\begin{figure}
\includegraphics[width=9cm]{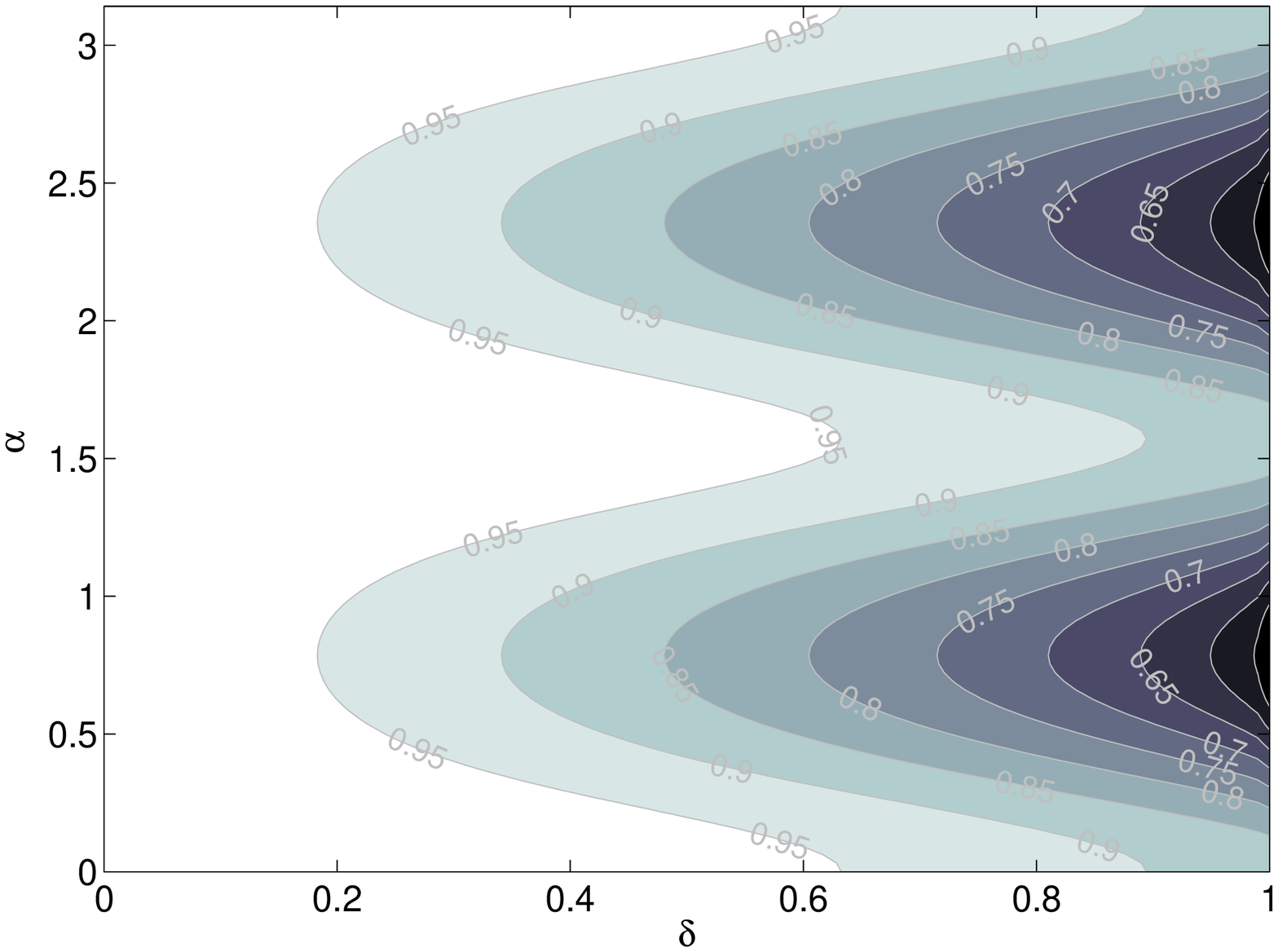}
\includegraphics[width=9cm]{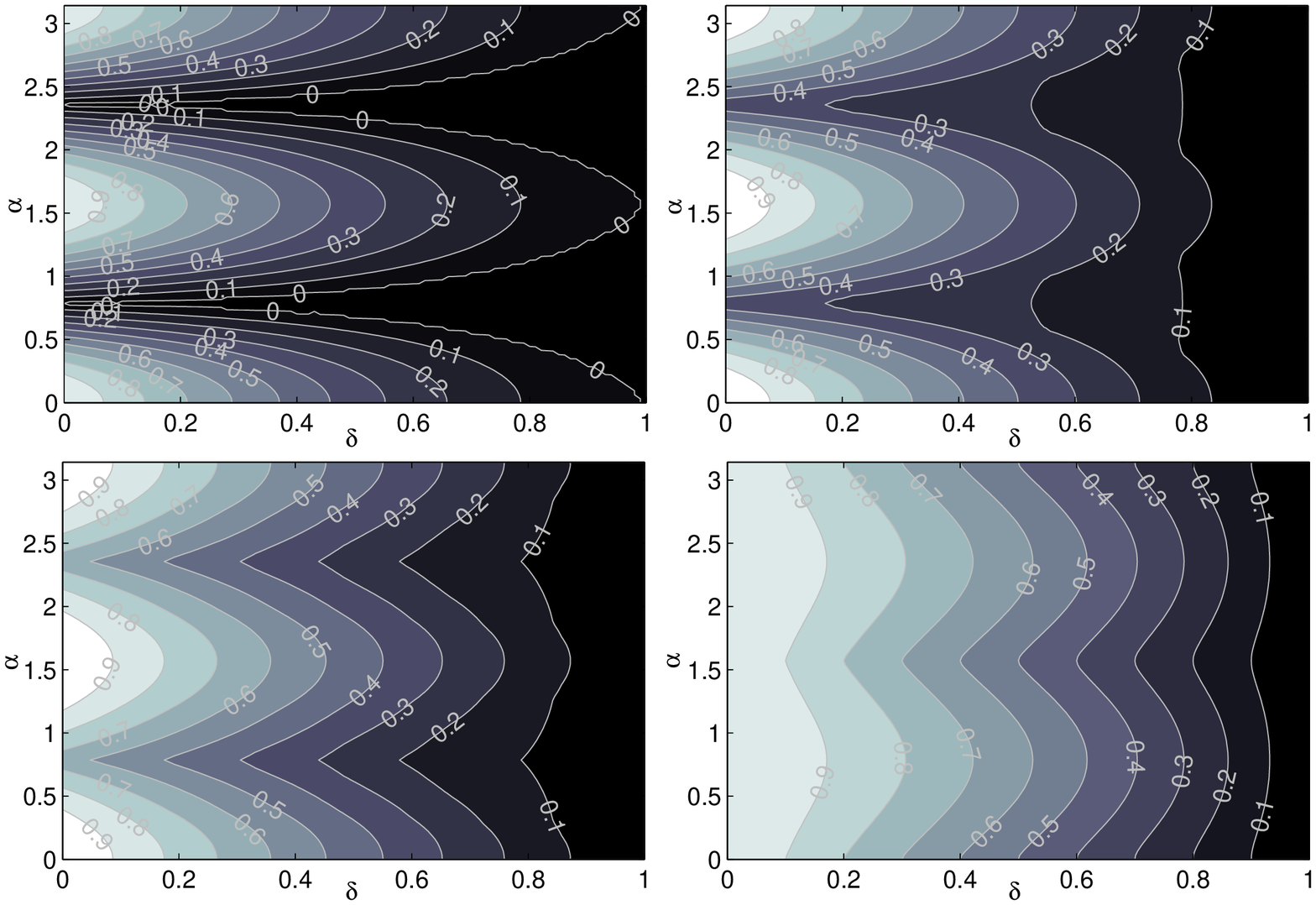}
\caption{\label{DephaseEnt} Fidelity and negativity as a function of dephasing strength, $\delta$, 
and initial state (paramaterized by $\alpha$ for $\beta = 0$): 
Top: Fidelity of unencoded qubit state after decoherence, syndrome measurement, and recovery operation.
Bottom: Entanglement measures for five qubit state before syndrome measurement: $N_1$ (upper-left), 
$N_{1,2}$ (upper-right), $N_{4,5}$ (lower-left), $N_2$ (lower-right). All entanglement measures except $N_2$ appear 
correlated with the decay of fidelity. The higher the amount of entanglement remaining between the qubits
the higher the fidelity of the stored information. Note also that $N_1$ is the only entanglement measure 
to exhibit ESD.  
}
\end{figure}
\end{center}

The effect of decoherence on the fidelity of the QEC code and the entanglement between the
physical qubits is shown in Fig. \ref{DephaseEnt}. Interestingly,
the robustness of the quantum information is very much dependent on the initial state. As an 
example, the decoherence strength at which the fidelity will fall below .95 varies 
widely, $.18 < \delta < .64$, depending on the initial state. The fidelity of initial states 
with low degrees of various types of entanglement decrease most quickly (states around 
$\alpha = \pi/4, \beta = 0$) while the fidelity of those states with the highest initial 
entanglement ($\alpha = 0,\pi/2, \beta = 0$) decreases most slowly. For stronger decoherence 
this difference in fidelity remains. Highly entangled initial states ($\alpha = 0,\pi/2$) 
retain a high level of fidelity $>.85$, even in the limit $\delta\rightarrow1$, but the fidelity 
of initial states with low entanglement decays all the way down to .5. Only one entanglement 
metric exhibits ESD: $N_1$. This implies that the first qubit is, in general, less integrated 
into the logical qubit than the other qubits. More relevant for this study is the demonstration 
that ESD plays little, if any, role in determining the success of the QEC code as clearly there 
are a host of entanglement metrics that do not decay to zero until $\delta\rightarrow1$.

While for most of the explored entanglement metrics the entanglement depends 
on the initial state, all states intially have the same $N_2$ entanglement. For this 
metric, states that are slower to lose this entanglement have a somewhat faster decrease 
in fidelity. From the above there does seem to be a correlation between the entanglement 
degradation (for all explored entanglement measures except $N_2$) and the fidelity of the 
stored quantum information. States with initially high degrees of entanglement lose their 
entanglement much more slowly than states with low initial entanglement and the fidelity 
of the quantum information for the former states remains much higher. 

Similar results to those of intial states $\beta = 0$ are found for states with $\beta = \pi/4$. 
However, the initial entanglement is generally slightly higher, and all states contain some 
$N_1$ entanglement. The $N_1$ entanglement does undergo ESD but at decoherence values $\delta \ge .8$.  

\subsection{Amplitude Damping}

We now turn to an independent qubit amplitude damping environment. As above, I explore the entanglement 
evolution of the qubits which make up the QEC code and compare it to the fidelity of the stored quantum 
information. The Kraus operators for this environment are:
\
\begin{equation}
K_1=\left(
\begin{array}{cc}
 1 & 0 \\
 0 & \sqrt{1-\delta}
\end{array}
\right); \;\;\;\; 
K_2=\left(
\begin{array}{cc}
 0 & \sqrt{\delta} \\
 0 & 0
\end{array}
\right)
\end{equation}
\
where the (time dependent) amplitude damping strength is denoted $\delta$. 

\begin{center}
\begin{figure}
\includegraphics[width=9.25cm]{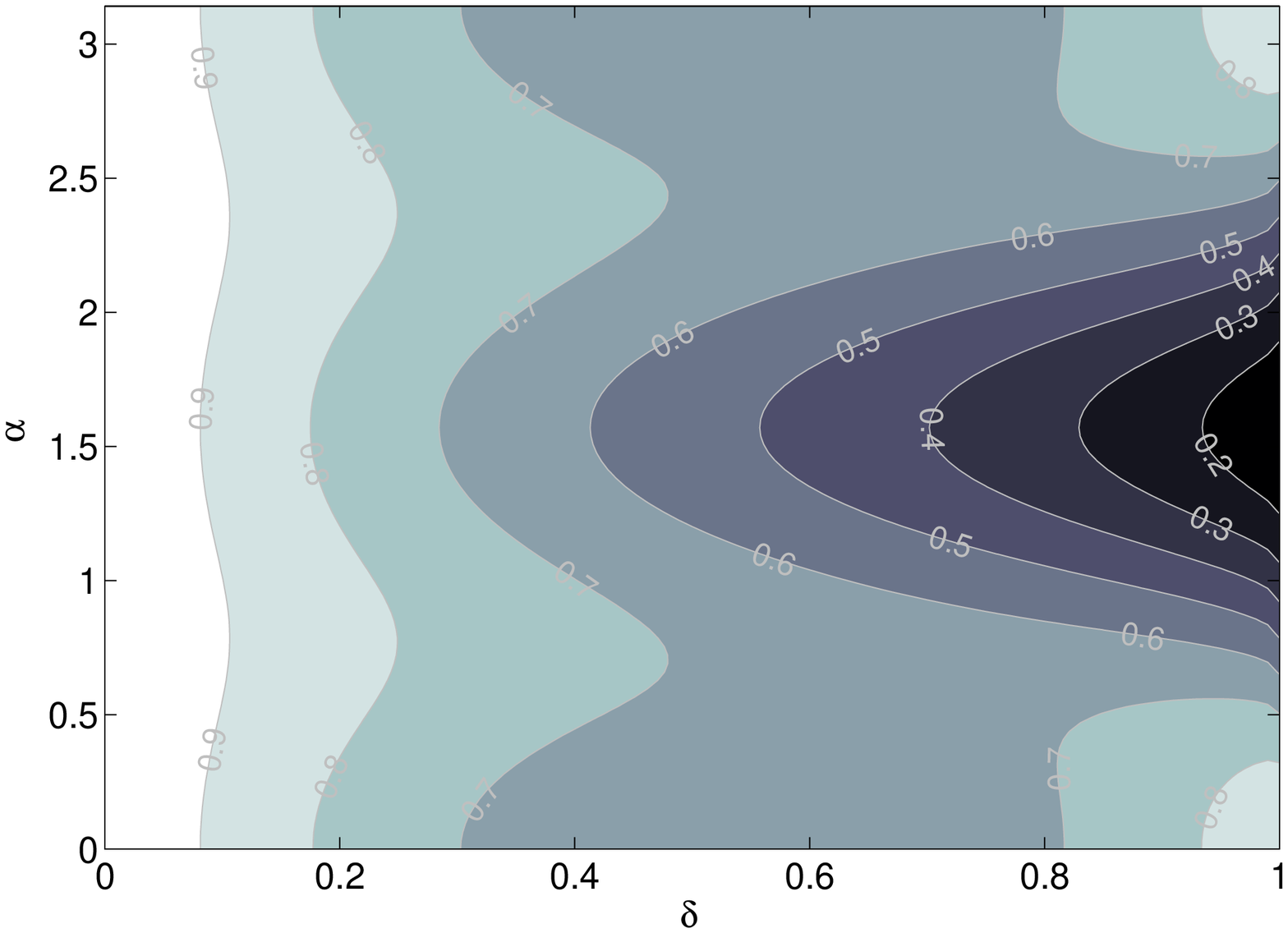}
\includegraphics[width=9cm]{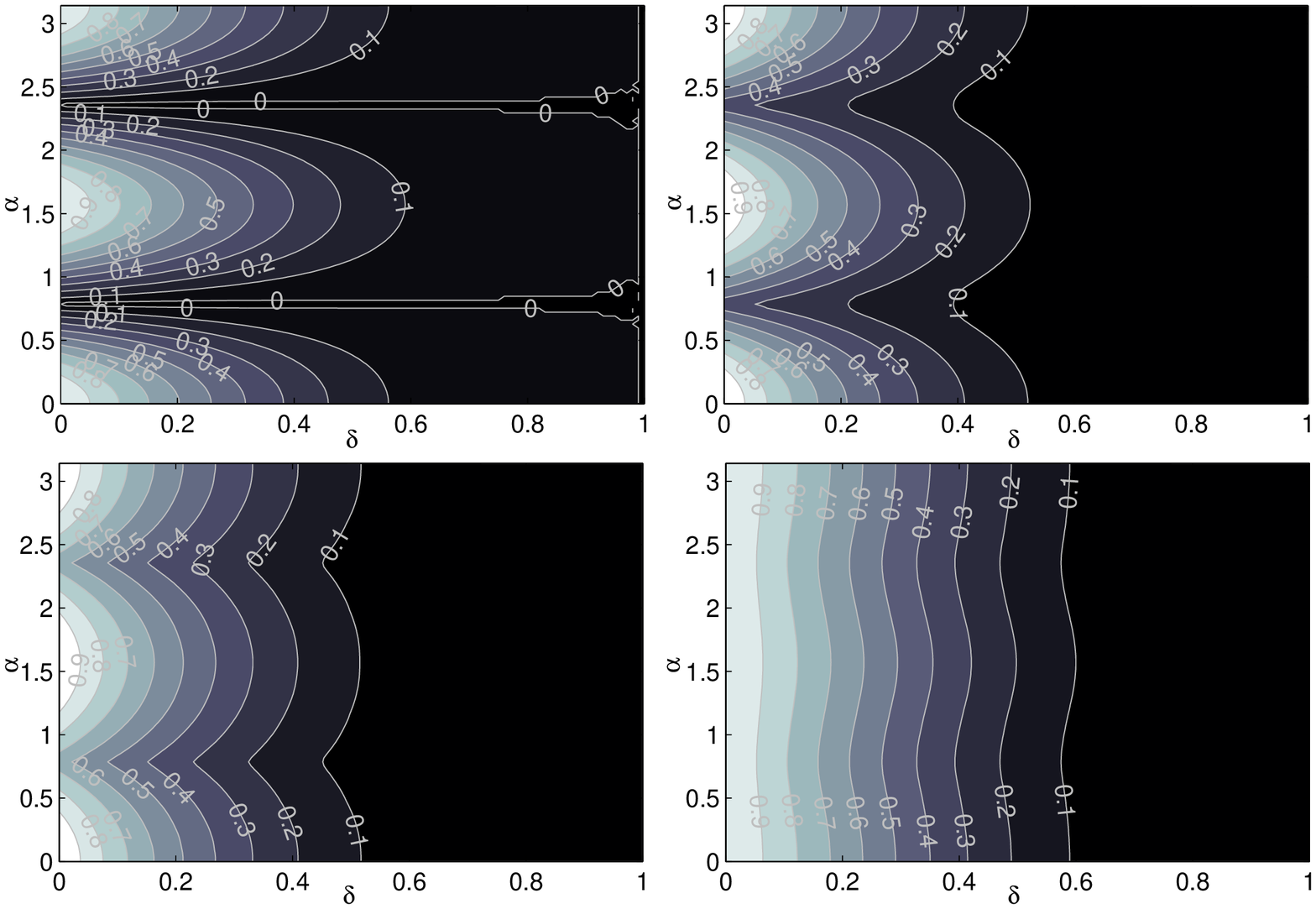}
\caption{\label{AmpEnt} Fidelity and negativity as a function of amplitude damping strength, $\delta$, 
and initial state (paramaterized by $\alpha$ for $\beta = 0$): 
Top: Fidelity of final qubit state after decoherence, syndrome measurement, and recovery operation.
Note that the fidelity behavior is much different than that of the phase damping environment and,
for certain states, increases with increasing $\delta$.
Bottom: Entanglement measures for five qubit state before syndrome measurement: $N_1$ (upper-left), 
$N_{1,2}$ (upper-right), $N_{4,5}$ (lower-left), $N_2$ (lower-right). Unlike the case of phase 
damping, the fidelity of the highly entangled initial states decays faster than for initial states
with less entanglement. For high values of decoherence the fidelity 
and entanglement behave in entirely different ways as explained in the text. Once again, $N_1$ is 
the only entanglement measure to exhibit ESD.  
}
\end{figure}
\end{center}

As seen in Fig.~\ref{AmpEnt}, the fidelity of the quantum information stored in the QEC code remains 
close to one under amplitude damping only for very low decoherence strengths. At higher 
decoherence strengths we find a remarkable difference of behavior between states close to $\alpha = 0$ 
and those close to $\alpha = \pi/2$. States close to the latter point exhibit a deep loss of 
fidelity, which falls below .2, much lower than the lowest fidelity exhibited in the phase damping 
environment. In contrast, states close to the former point exhibit the opposite behavior. After 
the expected decrease in fidelity due to the decoherence, the fidelity begins to {\it increase}. For 
$\alpha = 0$ the decoherence strength where the transition occurs is $\delta \simeq .56$. 
As $\alpha$ increases so does the $\delta$ where the transition occurs. The fidelity at $\alpha = 0, 
\delta = 1$ reaches the value of .875. Comparing this to the entanglement evolution we note that 
the initial entanglement and subsequent entanglement decay is the same for these two states. 
This clearly demonstrates the inability of entanglement degradation to indicate fidelity of the 
stored quantum information. 

In general, the entanglement evolution under amplitude damping reflects the opposite of what we found 
for the phase damping environment. In the amplitude damping environment it is the initial states with 
higher entanglement that experience a faster initial fidelity decay. In fact, the entanglement 
degradation under amplitude damping looks very similar to the entanglement degradation under 
phase damping, albeit the degradation occurs at a faster rate. The fidelity decay behavior, however,
is completely different. It is, for low $\delta$, much more uniform with respect to the initial state,
the fidelity increases as $\delta$ increases for certain initial states, and states around $\alpha = \pi/2$
which have the highest fidelity under phase damping, achieve the lowest fidelity under amplitude damping.
ESD is once again exhibited only for $N_1$ and only for very limited initial states, not the states that
achive the lowest fidelity. This demonstrates that ESD plays no role in determining the success of the QEC code.

\subsection{Depolarizing}

The final decohering environment we explore is an independent qubit depolarizing environment 
and, as above, I compare the entanglement evolution to the fidelity of the stored quantum 
information. The Kraus operators for this environment are:
\begin{equation}
K_1= \sqrt{1-\frac{3\delta}{4}}\sigma_0; \;\;\;\;
K_j = \frac{\sqrt{\delta}}{2}\sigma_j, 
\end{equation}
where $sigma_0$ is the identity and $\sigma_j$ are the Pauli spin operators,
$j = x,y,z$ and $\delta$ is now the (time dependent) depolarizing strength. 

In a depolarizing environment the fidelity drops approximately uniformly as a function of 
initial state before reaching $F(\alpha,\beta,\delta\rightarrow1) = .5$. 
In contrast, the entanglement decay behavior is similar to that of the
other environments in that the initial entanglement and the rate of decay depends on the 
initial state. $N_2$ however, does decay almost uniformly with initial state. In addition, 
the entanglement under depolarizing decays much more quickly than for the other decohering 
environments as does the fidelity. ESD is exhibited for all metrics 
and all initial states at $\delta \leq .5$, where the fidelity is 
$.55 < F(\alpha,0,\delta) < .6$.

\begin{center}
\begin{figure}
\includegraphics[width=9cm]{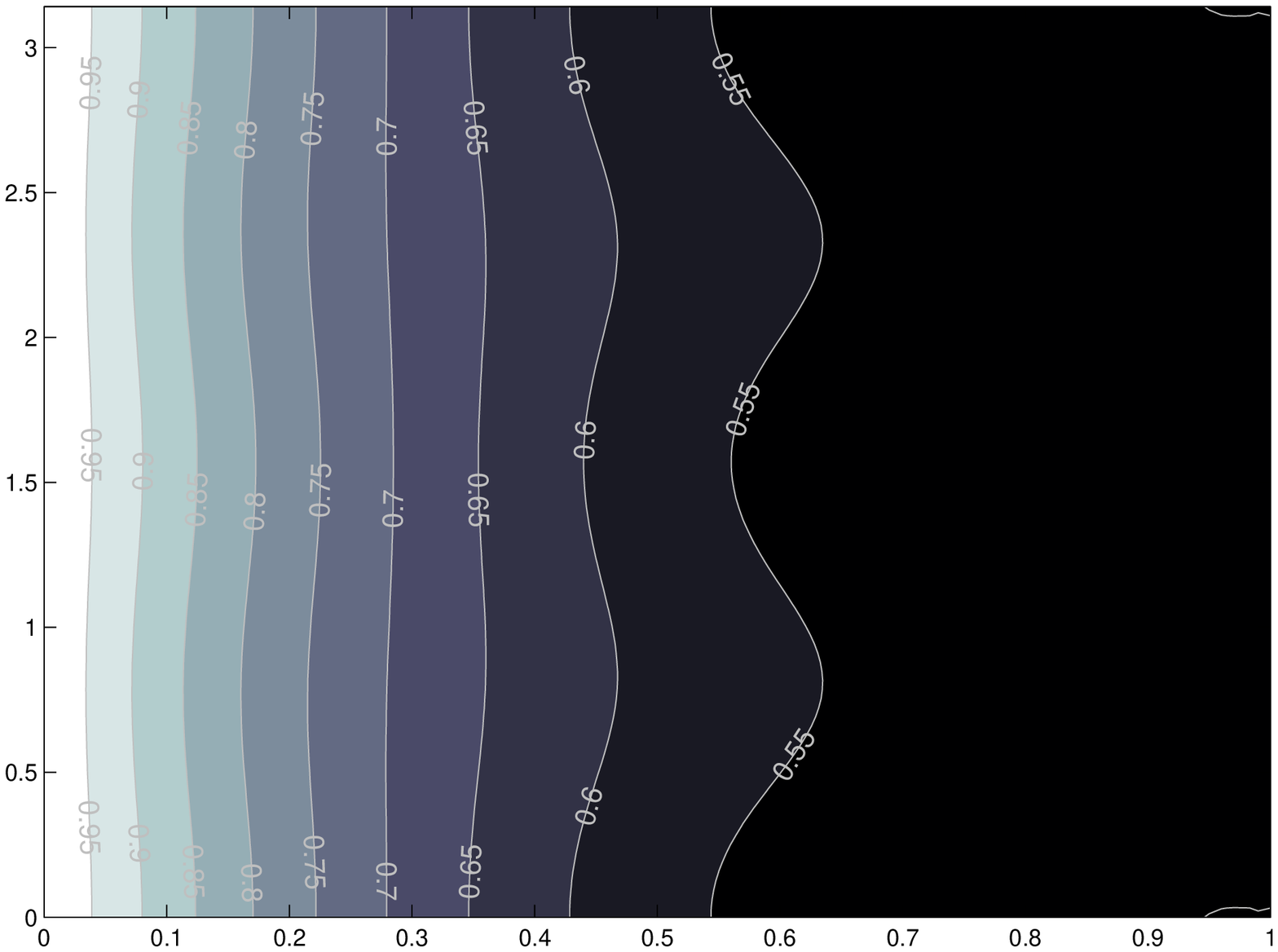}
\includegraphics[width=9cm]{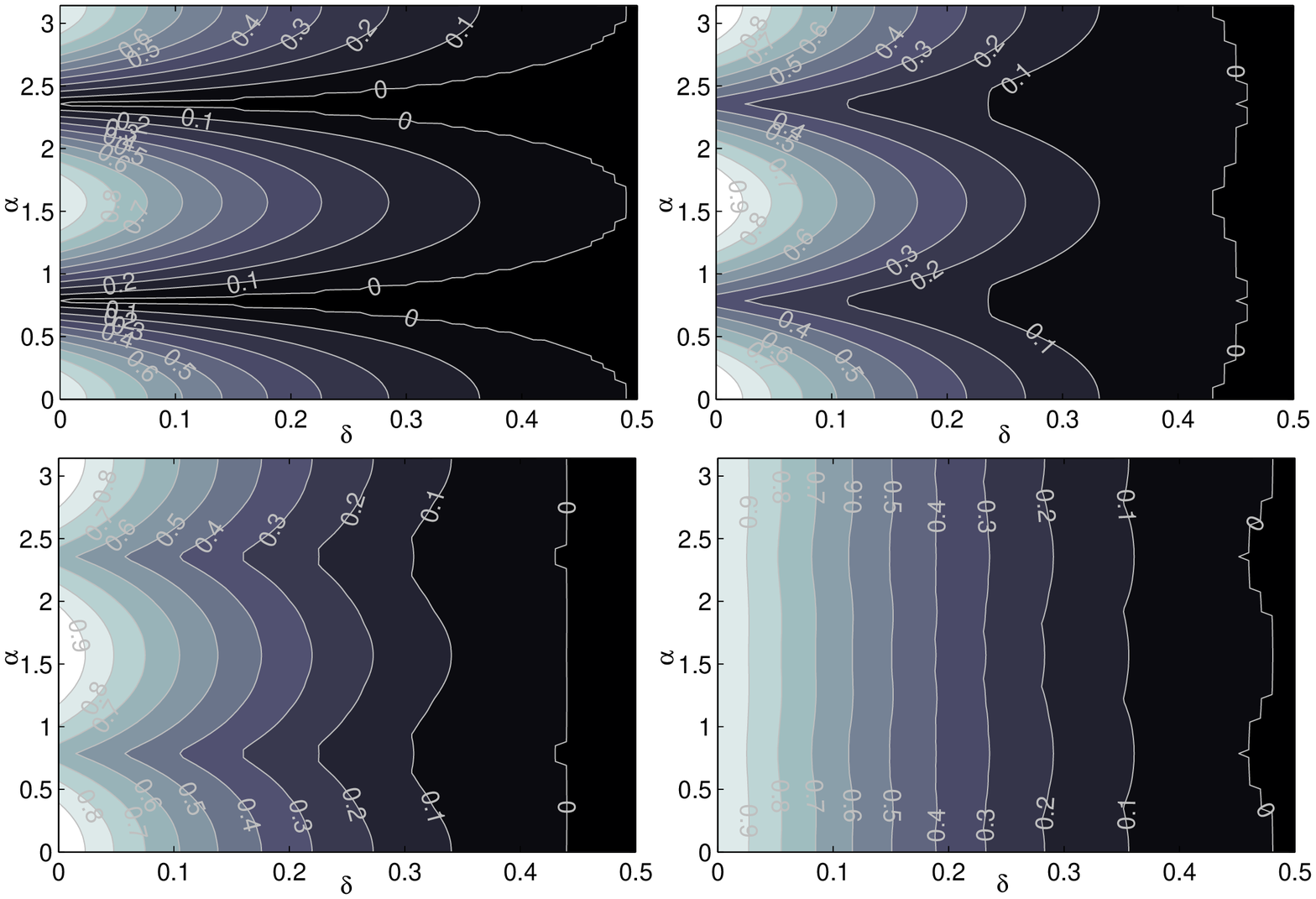}
\caption{\label{DepolEnt} Fidelity and negativity as a function of amplitude damping strength, 
$\delta$, and initial state (paramaterized by $\alpha$ for $\beta = 0$): 
Top: Fidelity of final qubit state after decoherence, syndrome measurement, and recovery operation.
For the depolarizing environment the fidelity decays almost uniformly with initial state.
Bottom: Entanglement measures for five qubit state before syndrome measurement: $N_1$ (upper-left), 
$N_{1,2}$ (upper-right), $N_{4,5}$ (lower-left), $N_2$ (lower-right). Note that we only show $0 < \delta < .5$. For 
$\delta > .5$ all entanglement measures are equal to zero, {\it i.e.} they have all undergone ESD.
The $N_2$ metric most closely mimics the fidelity decay.}
\end{figure}
\end{center}

\section{Discussion and Conclusions}

The goal of a QEC code is to store quantum information in 
such a way such that it remains unaffected by decoherence. When the decoherence 
affecting the physical qubits of the QEC is weak, the syndrome measurement will generally project the 
(unmeasured) qubits into an almost pure state which can be easily rotated to the nearly correct 
pre-decohered state. For QEC codes like the 7-qubit CSS code, the syndrome measurements are done 
on ancilla qubits and the projection `restores' mostly all entanglement that may have been 
destroyed by the decoherence. For stronger decoherence, quantum error correction will generally 
fail, meaning the syndrome measurement will project the qubits into a mostly incorrect state. 

For the five qubit code there is only one unmeasured qubit after syndrome measurement and thus no
entanglement remains. The above analysis, comparing the loss of entanglement to the fidelity of the stored
quantum information, was thus done utilizing entanglement metrics applied to the state before syndrome measurement. 
Study of other QEC codes will allow for comparisons between fidelity decay and entanglement remaining 
after syndrome measurements and may reveal a closer parallel between entanglement and fidelity.

For the five qubit QEC, the similarity of the entanglement degradation when the qubits are in either
an amplitude damping or phase damping environment, despite the complete lack of similarity for the
decay of stored quantum information fidelity, demonstrates that the entanglement decay is not a good 
indicator of fidelity. Furthermore, we saw that ESD is exhibited only when the qubits are placed in a 
depolarizing environment or for a few specific states in other decohering environments. Nevertheless,
fidelity under depolarizing decreases only to .5 while the fidelity in the other decohering environments 
in the limit of $\delta\rightarrow1$ may be higher or lower depending on the environment and the initial state.
Thus, we see that ESD has absolutely no effect on the overall success of the QEC code. 

Studies such as the one presented here allow us to frame, and partially answer, the question: what is the role 
of entanglement in quantum information processing. While the encoding of all states into a five qubit QEC 
contains entanglement any parallels between the entanglement evolution and fidelity appear superficial. 
This implies that the entanglement is present only because most states in Hilbert space happen to be 
entangled and it is the large size of Hilbert space that allows for the constuction of QEC codes.

\begin{acknowledgements}
It is a pleasure to thank G. Gilbert for helpful feedback 
and acknowledge support from the MITRE Technology Program under MIP grant \#20MSR053CA.
\end{acknowledgements}

\end{document}